%% file: Lattice04.tex
\documentclass[fleqn,twoside]{article}
\usepackage{espcrc2}
\usepackage{graphicx}

\newcommand{\AmS}{{\protect\the\textfont2
  A\kern-.1667em\lower.5ex\hbox{M}\kern-.125emS}}
\begin{document}

\title{QCD at Finite temperature and density with staggered and Wilson quarks\thanks{Supported 
by the Key NSF Project (10235040), and National and Guangdong
Ministries of Education.}
}
\author{Xiang-Qian Luo,%
\address[ZSU]{Department of Physics, Zhongshan University, 
Guangzhou 510275, China} and He-Sheng Chen$^{\mathrm{a}}$}

\begin{abstract}
One of the most challenging issues in particle physics is to study QCD in
extreme conditions. Precise determination of the QCD phase diagram on
temperature $T$ and chemical potential $\mu$ plane will provide valuable
information for quark-gluon plasma (QGP) and neutron star physics.
We present results for phase structure on the $(\mu, T)$ plane for
lattice QCD with Wilson fermions from strong coupling Hamiltonian analysis
and Kogut-Susskind Fermions from Lagrangian Monte Carlo
simulations at intermediate coupling. 
\end{abstract}

\maketitle


\section{Introduction}

For two massless quarks, several QCD inspired models 
suggest the existence of a tricritical point on the $(\mu,T)$ plane 
separating the first order transition line at lower $T$ and
larger $\mu$, and the second order transition line at higher $T$ and smaller $\mu$. 
Lattice gauge theory (LGT) should in principle give more reliable information on the QCD phase diagram.
Heavy-ion collision experiments may determine whether the theoretical prediction is correct. 

In Lagrangian formulation of SU(3) LGT,
the chemical potential $\mu$ is introduced by making the replacement:  
$U_{4}(x)=e^{\mu a}U_{4}(x)$, $U_{4}^{+}(x)=e^{-\mu a}U_{4}^{+}(x)$. 
It is well known that, for
real $\mu$, the effective fermionic action is complex and the standard MC technique
does not apply.  
The recent years have
seen enormous efforts\cite{Katz:2003up} on
solving this problem, and some very interesting information
\cite{D'Elia:2002gd,deForcrand:2002ci,Fodor:2002hs} 
on the phase diagram for QCD with KS fermions at large 
$T$ and small $\mu$ has been obtained from MC simulations. 
There is also a strong coupling analysis\cite{Azcoiti:2003eb}
of the color superconductivity phase at $T=0$.

Nevertheless, to precisely locate the
tricritical point and critical line at large $\mu$ is still an extremely
difficult task. 
QCD at large $\mu$ is of particular importance for neutron star physics. 
Hamiltonian formulation of LGT doesn't encounter the notorious
``complex action problem''. 
Recently, we proposed a Hamiltonian approach\cite{Gregory:1999pm,Luo:2000xi}
to LGT with naive fermions at finite $\mu$, and
extended it to Wilson fermions\cite{Fang:2002rk}. The chiral phase
transition at $T=0$ and some finite $\mu_C$ was found to be of first order.
In Ref. \cite{Bringoltz:2002qc}, the authors studied lattice QCD with SLAC fermions at $T=0$ 
in the strong coupling limit.

In this paper, we study the phase diagram on the $(\mu,T)$ plane by Lagrangian MC simulations with
Kogut-Susskind quarks and Hamiltonian strong coupling analysis with Wilson fermions.

\section{MC Simulation with $N_f=2$ and 4}
\label{our approach}

For small $\mu$, some approximation can be used in simulations. 
If $\mu$ is purely imaginary, the effective fermionic action is real
and the traditional MC methods still applies. Using Taylor
expansion, the physical observables at imaginary $\mu$ can be extrapolated to real $\mu$.

For imaginary $\mu=i\mu_I$, Roberge and Weiss\cite{Roberge:1986mm}
find that the partition
function  is periodic in  $\mu_I /T$ with periodicity
$2\pi /3$. The Polyakov loop is singular at $\mu_I /T=2\pi (k+1/2)/3$, with $k$ an integer.
In pure gauge theory, the partition function has the universal $Z(3)$ symmetry. In full QCD,
the fermionic determinant breaks the $Z(3)$ symmetry.
When $\mu_I$ is increased
from 0, there are phase transitions at
$\mu_I /T=2\pi (k+1/2)/3$. For $\mu_I /T<\pi /3$,
the simulation is physical and for $\mu_I /T>\pi /3$,
the result is only a simple copy of that for $\mu_I /T<\pi /3$, which is the artifact of imaginary $\mu$.
So this method  works only for $\mu_I /T<\pi /3$.

\begin{figure} [htbp]
\begin{center}
\includegraphics[totalheight=5in]{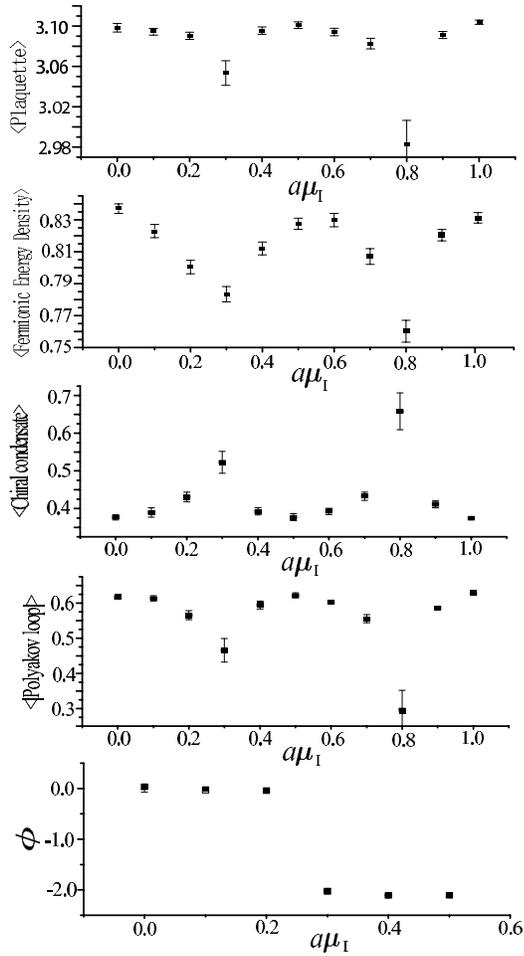}
\end{center}
\vspace{-1.0cm}
\caption{Results for KS quarks at $\beta =5.1$, $am=0.05$, and $N_{f}=4$. $\phi$ is the phase of Polyakov loop.}
\label{fig1}
\end{figure}

\begin{figure} [htbp]
\begin{center}
\vspace{-1.6cm}
\includegraphics[totalheight=4.0in]{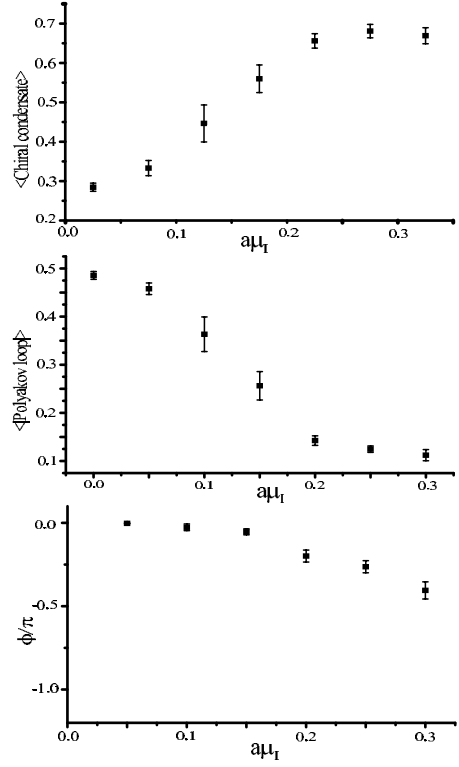}
\end{center}
\vspace{-1.0cm}
\caption{Results for KS quarks at $\beta =5.3$, $am=0.25$, and $N_{f}=2$.}
\label{fig2}
\end{figure}

We have simulated QCD with KS fermions using the standard
R and HMC algorithms, for $N_f=2$  and $N_f=4$ respectively.
We modified the MILC code to imaginary $i \mu_I$.
The molecular dynamics for each configuration consists of 20 steps with step size $\delta \tau$ is 0.02. 
The lattice size is $8^{3}\times 4$. The total trajectories are 4000. 
The measurements were done for every 40 trajectories.
Fig. \ref{fig1} and Fig. \ref{fig2} show our results.
The data for the chiral condensate, Polyakov loop and the its phase are consistent
with Refs. \cite{D'Elia:2002gd,deForcrand:2002ci}. 
In addition, we measured the plaquette and fermionic energy density.

\section{Tricritical Point at Strong Coupling}
\label{critical}

In the Hamiltonian formulation, one can study QCD at real $\mu$.
At strong coupling, the effective Hamiltonian
with four fermion interactions\cite{Fang:2002rk,Fang:2001ry} is the result from
integrating out the gauge fields.
For convenience, we rescale the chemical potential and temperature as
$\mu'=\mu /(3K /a)$ and $T'=T/(3K/a)$, with $K$ being the effective coupling of
four fermion interactions.
At the second order phase transition where the chiral condensate and the
dynamical mass of quark vanish continuously, there is only one global
minimum in the grand thermodynamic potential. 
For Wilson fermions and $N_f/N_c<1$, where $N_c=3$ is the number of colors, 
we obtain an equation for the critical line 
\begin{eqnarray}
\mu_C^{\prime}&=& \left( 1+r^2 \right) \sqrt{ 1-{\frac{2T_C^{\prime}}{1+3r^2}
}}  \nonumber \\
&+&T_C^{\prime}\ln {\frac{ 1+\sqrt{ 1-{\frac{2T_C^{\prime}}{1+3r^2}}} }{1-
\sqrt{ 1-{\frac{2T_C^{\prime}}{1+3r^2}}}}} ,  
\label{second_order}
\end{eqnarray}
which is depicted by the dotted line for $r=1$ in Fig. \ref{fig3}. 

Below some finite $T^{\prime}_3$,  there is a
first order chiral phase transition line 
\begin{eqnarray}
\mu^{\prime}_C =1+2r^2 .  \label{firstorder}
\end{eqnarray}
from some finite $T^{\prime}_3$ down to $T^{\prime}=0$. this is illustrated
by the solid line for $r=1$ in Fig. \ref{fig3}.

The point when two lines described by Eq. (\ref{second_order}) and 
Eq. (\ref{firstorder}) join at lower $T^{\prime}$ is the tricritical point 
$(\mu^{\prime}_3,T^{\prime}_3)$. For $r=1$, we find 
$(\mu^{\prime}_3,T^{\prime}_3)=(3, 0.4498)$, i.e. the circle in Fig. \ref{fig3}. The phase
structure for any $r\ne 0$ is qualitatively the same. Details
can be found in Ref. \cite{Luo:2004mc}.

\section{Summary}
\label{Symmary}

In the preceding sections, 
we present some preliminary MC data for QCD with KS fermions at some finite $\beta$  and imaginary $\mu$.
These can be converted to some $T$ and real $\mu$. 
The results for $N_f=2$ might indicate the existence of a second order chiral phase transition, if the data are
extrapolated to the chiral limit.
Those for $N_f=4$ indicate the existence of a first order phase transition.
To determine whether a tricritical point exists for $N_f=2$, further detailed study is required.

We also investigate the QCD phase diagram in
Hamiltonian lattice formulation with Wilson fermions. At the strong
coupling, we find a tricritical point on the $(\mu,T)$ plane, which has not
been found in previous work in the Hamiltonian formalism with KS or naive fermions. 
Our findings imply that on the $(\mu,T)$ plane, at least in the Hamiltonian formulation at the strong coupling,
the phase structure of QCD with Wilson fermions (without species doubling)
might be qualitatively different from naive or KS fermions (with
species doubling).

\vspace{-5.0cm}

\begin{figure}[htb]
\vspace{-1.7cm}
\input Phase.tex
\vspace{-1.0cm}
\caption{Phase diagram for Wilson quarks. The solid and dotted lines stand respectively for
the 1st and 2nd order transitions. The circle is the tricritical point.}
\label{fig3}
\end{figure}

\end{document}

%% file: Phase.tex
\begingroup%
  \makeatletter%
  \newcommand{\GNUPLOTspecial}{%
    \@sanitize\catcode`\%=14\relax\special}%
  \setlength{\unitlength}{0.1bp}%
\begin{picture}(2087,1511)(0,0)%
\special{psfile=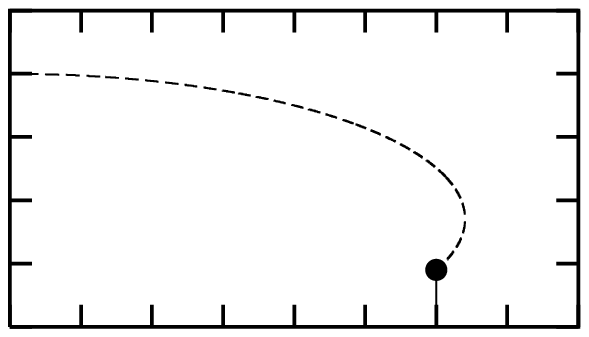 llx=0 lly=0 urx=418 ury=353 rwi=4180}
\put(1218,1361){\makebox(0,0){$r=1$}}%
\put(1218,50){\makebox(0,0){$\mu'$}}%
\put(100,755){%
\special{ps: gsave currentpoint currentpoint translate
270 rotate neg exch neg exch translate}%
\makebox(0,0)[b]{\shortstack{$T'$}}%
\special{ps: currentpoint grestore moveto}%
}%
\put(2037,200){\makebox(0,0){4}}%
\put(1832,200){\makebox(0,0){3.5}}%
\put(1628,200){\makebox(0,0){3}}%
\put(1423,200){\makebox(0,0){2.5}}%
\put(1219,200){\makebox(0,0){2}}%
\put(1014,200){\makebox(0,0){1.5}}%
\put(809,200){\makebox(0,0){1}}%
\put(605,200){\makebox(0,0){0.5}}%
\put(400,200){\makebox(0,0){0}}%
\put(350,1211){\makebox(0,0)[r]{2.5}}%
\put(350,1029){\makebox(0,0)[r]{2}}%
\put(350,847){\makebox(0,0)[r]{1.5}}%
\put(350,664){\makebox(0,0)[r]{1}}%
\put(350,482){\makebox(0,0)[r]{0.5}}%
\put(350,300){\makebox(0,0)[r]{0}}%
\end{picture}%
\endgroup
 